\documentclass[twocolumn,showpacs,preprintnumbers,amsmath,amssymb]{revtex4}

\usepackage{epsfig}
\usepackage{graphicx}
\usepackage{dcolumn}
\usepackage{bm}

\begin{document}

\preprint{Draft Revision 1.2}

\title{An efficient quantum algorithm for the one-dimensional Burgers equation}

\author{Jeffrey Yepez}%
\email{Jeffrey.Yepez@hanscom.af.mil}
\homepage{http://qubit.plh.af.mil}
\affiliation{
Air Force Research Laboratory \\
29 Randolph Road, Hanscom Field, Massachusetts 01731}

\date{August 14, 2002}

\begin{abstract}
We analyze one-dimensional classical and quantum microscopic lattice-gas models governed by a lattice Boltzmann equation at the mesoscopic scale, achieved by ensemble averaging over microscopic realizations.  The models are governed by the Burgers equation at the macroscopic scale, achieved by taking the limit where the grid size and time step both approach zero and by performing a perturbative Chapman-Enskog  expansion.  The quantum algorithm exploiting superposition and entanglement is more efficient than the classical one because the quantum algorithm requires less memory. Furthermore, its viscosity can be made arbitrarily small. 
\end{abstract}

\maketitle


Here we present the simplest example where a quantum computer is demonstrably more efficient at numerically predicting the time-dependent solutions of an important nonlinear one-dimensional partial differential equation,  the classical Burgers equation
\begin{equation}
\label{speedup-burgers-equation}
\partial_t u +  u\partial_x u = \nu  \partial_x^2 u,
\end{equation}
a simplified model of turbulence and shock formation with flow field $u(x,t)$, sound speed $c_s$, and kinematic viscosity $\nu$. We shall derive (\ref{speedup-burgers-equation}) as the general effective field theory describing the  large-scale behavior of microscopic one-dimensional lattice gas models with two particles per site.  The occupation probabilites for the two particles at position $x$ and at time $t$ are denoted by $p_+(x,t)$ and $p_-(x,t)$, respectively.   The mesoscopic kinetic transport dynamics is governed by the lattice Boltzmann equation
\begin{equation}
\label{speedup-mesoscopic-transport-equation}
p_\pm(x\pm\delta x, t+\delta t) = p_\pm(x,t) \pm\Omega(p_+,p_-),
\end{equation}
where $\Omega$ denotes the collision term, a nonlinear function of $p_+$ and $p_-$.  The particular functional form of $\Omega$ depends on the model type, either classical or quantum mechanical.  For any $\Omega$, the model (\ref{speedup-mesoscopic-transport-equation}) always conserves particle number density defined as $\rho \equiv p_+ + p_-$.  
The propagation speed of particles is the ratio of the lattice cell size to the time step interval, $c=\frac{\delta x}{\delta t}$.   In these models, the macroscopic flow field is
\begin{equation}
\label{speedup-flow-field}
u(x,t) = c\left(\rho(x,t)-1\right). 
\end{equation}
%


The total classical computational complexity is the product of the lattice size $N_x$,  the ensemble of size $N$, and local resources of size $\varrho$
\begin{equation}
\label{speedup-classical-complexity}
C_{\hbox{\tiny cl}} \equiv N N_x \varrho.
\end{equation}
The quantity $\varrho$ is the amount of resources needed to encode $\Omega$ for the local microscopic collisions.
Since we are modeling the time evolution of a classical system, the value of $u(x,t)$ must be known everywhere at every time step whence the need for continual measurement in any quantum model. Therefore, the memory load factor for the lattice is still $N_x$ and not $\log_2 N_x$ in the quantum case \cite{yepez-ijmpc00a}.   Furthermore, because of wave function collapse by Von Neuman projective measurement, ensemble averaging is required just as in the classical case, so  a lower bound for the quantum computational complexity $C_{\hbox{\tiny qu}}$ is
\begin{equation}
\label{speedup-quantum-complexity}
C_{\hbox{\tiny qu}} \ge N N_x \log_2 \varrho.
\end{equation}
For some quantum computing technologies, such as spatial nuclear magnetic resonance quantum computing \cite{yepez-pravia-cpc2001, yepez-pravia-pre2002}, $N$ may be counted as order unity overhead.   Relating $\varrho$ to the minimum number of bits per node needed to encode (\ref{speedup-mesoscopic-transport-equation}), as a parsimonious demonstration of (\ref{speedup-classical-complexity}) and (\ref{speedup-quantum-complexity}),  it is known (\ref{speedup-burgers-equation}) can be modeled classically with 3 bits per node whereas we prove quantum mechanically only 2 qubits per node is required.


Multiplying (\ref{speedup-burgers-equation}) by $u$ and integrating over all space, with periodic boundaries, gives a relation for energy conservation
where the time-rate of change of the turbulent kinetic energy density $\partial_t(\frac{u^2}{2})$ is balanced by the viscous dissipation $\varepsilon \equiv \nu \left(\frac{\partial u}{\partial x}\right)^2 \sim \frac{u_{\cal L}^3}{\cal L}$, where  $\cal L$ is the characteristic scale of the largest feature in the flow field and $u_{\cal L}$ is the standard deviation of the turbulent kinetic energy or eddy velocity at that scale.

The flow velocity, the kinematic viscosity, and the viscous dissipation quantities have the dimensions: $\left[ u \right] =  \frac{L}{T}$, $\left[\nu \right]  =  \frac{L^2}{T}$, and $\left[\varepsilon \right] = \frac{L^2}{T^3}$.
The dissipation scale $\lambda\equiv  \left(\frac{\nu^3}{\varepsilon}\right)^{\frac{1}{4}}$ is the smallest spatial scale where macroscopic effective field theory (\ref{speedup-burgers-equation}) is physically applicable and the smallest physical velocity at the dissipation scale is the {\it dissipation-scale velocity} $u_\lambda  \equiv  \left(\nu \varepsilon\right)^{\frac{1}{4}}=\frac{\nu}{\lambda}$. 

The Reynolds number characterizing the fluid's nonlinearity is $\hbox{Re} \equiv \frac{\cal L}{\lambda } \frac{u_{\cal L}}{u_\lambda}= \frac{ {\cal L} u_{\cal L}}{\nu}$, and using $\varepsilon=\frac{u_{\cal L}^3}{\cal L}$ to eliminate $\cal L$, we also have $\hbox{Re} = \frac{ u_{\cal L}^4 }{\nu\varepsilon} = \left(\frac{u_{\cal L}^2 }{u_\lambda^2}\right)^2= \left(\frac{\cal L}{\lambda }\right)^{\frac{4}{3}} $.  The computational complexity of numerically modeling fluidic behavior at the macroscopic scale can be expressed as a function of Re.  First, the number of grid points $N_x$  to sufficiently resolve the flow field down to the dissipation scale is $N_x = \frac{\cal L}{\lambda}= \hbox{Re}^{\frac{3}{4}}$.
Second, we get $N$ by physically limiting $|\delta p_\pm|$.  
Since, the occupation probabilities are measured by ensemble averaging over $N$ independent microscopic realizations, $|\delta p_\pm| \simeq  \frac{1}{\sqrt{N}}$, 
due to  either classically stochastic shot noise or quantum mechanically stochastic projective measurement.
Following Orszag and Yakhot \cite{orszag-86}, the value of the statistical fluctuation  $\delta u(x,t)$ of the numerical flow field must be much much less than $u_\lambda(x,t)$ of the macroscopic effective field theory (\ref{speedup-burgers-equation}).  Using (\ref{speedup-flow-field}), we have $u_\lambda  \gg c (\delta p_+ + \delta p_-)\sim \frac{c}{\sqrt{N}}$.  Hence, the ensemble size $N \gg \frac{c^2}{u_\lambda^2}\sim\frac{ \hbox{Re}^{\frac{1}{2}}}{M^2}$, where the {\it Mach number} is $M\equiv u_{\cal L}/c$.


It is convenient to treat the occupation probabilities as a two-component field
\begin{equation}
|p\rangle = \left(\begin{matrix}p_+ \cr p_-\end{matrix}\right).
\end{equation}
We expand $|p\rangle$ about equilibrium its value denoted $|d\rangle$ so that $|p\rangle = |d\rangle +  |\delta p\rangle +{\cal O}(\varepsilon^2)$, where $\varepsilon\sim\frac{1}{N_x}$ is the Knudsen number.
The equilibrium condition $\left.\Omega\right|_{p=d}=0$ leads to a tractable polynomial equation for $d_\pm$, whence the linearized finite-difference equation
\begin{equation}
\label{speedup-mesoscopic-transport-equation-linearized}
|p(x\pm\delta x, t+\delta t)\rangle - |p(x,t)\rangle = J |\delta p(x,t)\rangle,
\end{equation}
where the Jacobian of the collision term is
\begin{equation}
J \equiv 
\left.\left(\begin{matrix}
 \frac{\partial\Omega}{\partial p_+} &  \frac{\partial\Omega}{\partial p_-} \cr
 -\frac{\partial\Omega}{\partial p_+} &  -\frac{\partial\Omega}{\partial p_-} \cr
\end{matrix}\right)\right|_{p=d}
=
\left(\begin{matrix}
 J_+ & J_- \cr
- J_+ & -J_- \cr
\end{matrix}\right).
\end{equation}
The left and right eigenvectors of $J$ are
\begin{eqnarray}
\langle\xi_1|  = 
\left(\begin{matrix}
1 &
1
\end{matrix}\right) \hspace{0.25in}
&
|\xi_1\rangle  = &
\frac{1}{J_- - J_+}\left(\begin{matrix}
 J_- \cr 
-J_+
\end{matrix}\right) \hspace{0.25in} 
 \\
\langle\xi_2|  = 
\frac{1}{J_+ - J_-}\left(\begin{matrix}
J_+ & 
J_-
\end{matrix}\right) \hspace{0.25in}
&
|\xi_2\rangle  = & 
\left(\begin{matrix}
 1\cr
-1\cr
\end{matrix}\right) \hspace{0.25in}
\end{eqnarray}
with associated eigenvalues $\lambda_1 = 0$ and $\lambda_2 = J_+ - J_-$.
$\langle \xi_i | \xi_j\rangle = \delta_{ij}$.
$J$ may be rewritten as
\begin{equation}
J = \lambda_2 |\xi_2\rangle\langle\xi_2|
=\
\left(\begin{matrix}
 J_+ & J_- \cr
-J_+ & -J_- \cr
\end{matrix}\right)
\end{equation}
$J$ is singular.  Nevertheless, its generalized inverse is
\begin{equation}
\label{speedup-generalized-inverse}
J^{-1}_{\hbox{\tiny gen}} = \frac{1}{\lambda_2} |\xi_2\rangle\langle\xi_2|
=\frac{1}{J_+ -J_-}
\left(\begin{matrix}
 J_+ & J_- \cr
-J_+ & -J_- \cr
\end{matrix}\right).
\end{equation}

Now we invoke the continuum limit where $\delta x\rightarrow 0$ and $\delta t\rightarrow 0$ so $|p\rangle$ is a continuous and differentiable two-component field.
We obtain a first order equation by Taylor expanding (\ref{speedup-mesoscopic-transport-equation-linearized}) in $x$ and $t$ and keeping terms only terms first order in $\varepsilon$:
\begin{equation}
\sigma_z\delta x\partial_x | d\rangle = J | \delta p\rangle+{\cal O}(\varepsilon^2),
\end{equation}
where $\sigma_z =\left(\begin{matrix}1&0\cr 0& -1\end{matrix}\right)$.  Then using (\ref{speedup-generalized-inverse}) we have
\begin{equation}
| \delta p\rangle = \frac{1}{J_+-J_-}\sigma_z\delta x\partial_x | d\rangle+{\cal O}(\varepsilon^2) .
\end{equation}
Taking the difference of the respective components gives
\begin{equation}
\label{speedup-1st-order-solution}
\delta p_+-\delta p_- = \frac{1}{J_+-J_-}\delta x \partial_x\rho +{\cal O}(\varepsilon^2).
\end{equation}
Similarly from (\ref{speedup-mesoscopic-transport-equation-linearized}), we obtain the second order equation:
\begin{eqnarray}
\nonumber
\delta t\partial_t| d\rangle&+&
\sigma_z\delta x\partial_x\left( | d\rangle+ | \delta p\rangle\right)\\
&+&\frac{\delta x^2}{2}\partial_x^2 | d\rangle
+{\cal O}(\varepsilon^3) = 
\left(
\begin{matrix}
\Omega\cr
-\Omega
\end{matrix}
\right)
.
\end{eqnarray}
We now take the sum of the respective components:
\begin{eqnarray}
\nonumber
\delta t\partial_t\rho&+&\delta x\partial_x\left(  d_+- d_-+ \delta p_+-\delta p_-\right)\\
&+&\frac{\delta x^2}{2}\partial_x^2\rho
+{\cal O}(\varepsilon^3) = 
0.
\end{eqnarray}
Inserting (\ref{speedup-1st-order-solution}) into the above equation gives the general effective field theory for any one-dimensional two-particle-per-site lattice gas conserving particle number
\begin{eqnarray}
\label{speedup-effective-field-theory}
\nonumber
\partial_t\rho &+& c\partial_x\left(  d_+- d_-\right)+\frac{\delta x^2}{\delta t}\frac{\partial_x(J_+-J_-)}{(J_+-J_-)^2}\partial_x\rho\\
&+&\frac{\delta x^2}{2\delta t}\left(\frac{2}{J_+-J_-}+1\right)\partial_x^2\rho+{\cal O}(\varepsilon^3) = 
0.
\end{eqnarray}
%

The Brieger-Bonomi model of the Burgers equation has one bit per site with a 3-bit stencil where the center bit is updated by a stochastic Masters equation \cite{brieger-jsp92}.  This nonlocal classical model is equivalent to a 3-bit per site local lattice-gas model. We briefly discuss the classical Boghosian-Levermore lattice-gas model \cite{boghosian-87,boghosian-89} before proceeding to our quantum model. The classical collision term is
\begin{equation}
\Omega = \frac{1}{2}(p_- - p_+)+\frac{\alpha}{2}\left(p_+ + p_- - 2p_+p_-\right),
\end{equation}
where $\alpha$ is occupation probability of an additional random bit used to bias the collision \footnote{Actually, in the Boghosian-Levermore model, 59 additional bits per site were used to generate a single random bit in a strictly reversible fashion.}. 
Taking the equilibrium to be of the form $d_\pm = \frac{\rho}{2}\pm A$,
then $\left.\Omega\right|_{p=d}=0$ leads to a quadratic equation in $A$.  We take the negative root $A=\alpha \frac{\rho}{2}\left(1-\frac{\rho}{2}\right)$, and hence
\begin{equation}
\label{speedup-peqdiff-burgers}
d_+-d_- = \alpha\rho(1-\frac{\rho}{2}).
\end{equation}
With $J_\pm = \left.\frac{\partial\Omega}{\partial p_\pm}\right|_{p=d}$, we find
\begin{equation}
\label{speedup-Jdiff-burgers}
J_+-J_- = -1 + 2\alpha^2 \rho(2-\rho)\simeq -1 + {\cal O}(\alpha^2).
\end{equation}
Then substituting the two results (\ref{speedup-peqdiff-burgers}) and (\ref{speedup-Jdiff-burgers}) into (\ref{speedup-effective-field-theory}) yields the desired effective field theory
\begin{equation}
\label{speedup-burgers-effective-field-theory}
\partial_t\rho+c\alpha(1-\rho)\partial_x\rho-\frac{\delta x^2}{2\delta t}\partial_x^2\rho+{\cal O}(\varepsilon^3,\varepsilon\alpha^2)= 
0,
\end{equation}
which is the nonlinear Burgers equation for $u=c(\rho - 1)$ with variable sound speed $c_s = c \alpha$ and fixed kinematic viscosity $\nu = \frac{\delta x^2}{2\delta t}$.

\begin{figure}[htbp]
\begin{center}
\epsfxsize=2.5in
\epsffile{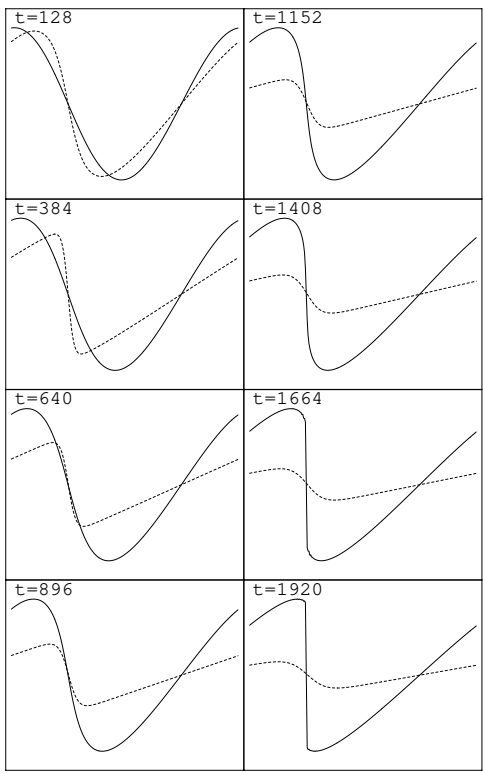}
\end{center}
\caption{\footnotesize The  2-qubit quantum algorithm with $\theta=\frac{\pi}{4}$ and $\zeta=\xi=0$ (solid curve), 3-bit Boghosian-Levermore classical algorithm with $\alpha=.707$ (dashed curve),  and the 2-bit classical algorithm (dotted curve) with $\alpha=1$.  There is good agreement between the quantum algorithm and the Boghosian-Levermore algorithm in this case for equal kinematic viscosities $\nu=\frac{\delta x^2}{2\delta t}$.  However, the 2-bit classical algorithm does not model the Burgers equation demonstrating that at least 3-bits per node are required in the classical case.  Hence the quantum algorithm requires less memory to achieve the same result.}
\label{ViscousQuantumVs3BitVs2Bit}
\end{figure}

In the special case when $\alpha =0$, (\ref{speedup-burgers-effective-field-theory}) reduces to the diffusion equation.  However, there exists an abnormal case when $\alpha =1$ where the random bit is fixed to the value of 1. The  3-bit model reduces to a 2-bit one that does not model the Burgers equation.  Its collision term is $\Omega = (1-p_+)p_-$ and the equilibrium occupations are $d_+ = \rho$ and $d_- = 0$.  The components of the Jacobian matrix are $J_+ = 0$ and $J_- = 1+\rho$. 
Hence $d_+-d_-=\rho$ and $J_+ - J_- = -1 - \rho$, so (\ref{speedup-effective-field-theory}) becomes
\begin{equation}
\label{speedup-diffusive-effective-field-theory}
\partial_t\rho+
c\partial_x\rho
+\frac{\delta x^2}{\delta t}\frac{\left(\partial_x\rho\right)^2}{(1-\rho)^2}
+\frac{\delta x^2}{\delta t}\left(\frac{\rho+1}{\rho-1}\right)\partial_x^2\rho+{\cal O}(\varepsilon^3) = 
0.
\end{equation}
See figure~\ref{ViscousQuantumVs3BitVs2Bit} for numerical solutions of (\ref{speedup-burgers-effective-field-theory}) and (\ref{speedup-diffusive-effective-field-theory}).


%
\begin{figure}[htbp]
\begin{center}
\epsfxsize=2.5in
\epsffile{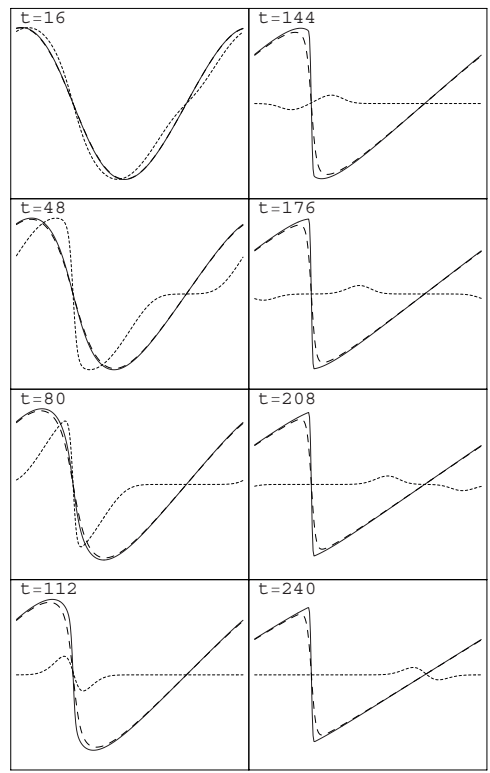}
\end{center}
\caption{\footnotesize The 2-qubit quantum algorithm with $\theta=1.5$ radians and $\zeta=\xi=0$ (solid curve) and the 3-bit Boghosian-Levermore classical algorithm with $\alpha=0.5$ (dotted curve).  This demonstrates that the quantum algorithm can model a low viscosity fluid when $\theta\simeq \frac{\pi}{2}$ and is therefore computationally efficient.}
\label{QuantumVsViscousClassical}
\end{figure}

Now we consider the quantum algorithm for the Burgers equation.
Initially we encode the qubits with their respective occupation probabilities
\begin{equation}
|q_\pm\rangle = \sqrt{p_\pm}|1\rangle + \sqrt{1-p_\pm}|0\rangle,
\end{equation}
ignoring the possibility of an internal phase angle.
The initial ket $|\psi\rangle = |q_+\rangle \otimes |q_-\rangle= \sqrt{p_+ p_-}|11\rangle + \sqrt{p_+(1-p_-)}|10\rangle+ \sqrt{(1-p_+)p_-}|01\rangle  + \sqrt{(1-p_+)(1-p_-)}|00\rangle$ is transformed by application of a unitary matrix:
\begin{equation}
\label{speedup-update-equation}
|\psi'\rangle = \hat U |\psi\rangle.
\end{equation}
The identity matrix and $\hat n = \left(\begin{matrix} 1 & 0 \cr 0 & 0 \end{matrix}\right)$ denote the single qubit number operator,
the multi-qubit number operators $\hat n_1 = \hat n \otimes {\bf 1}$ and   $\hat n_2 ={\bf 1}\otimes \hat n$ are used to determine the new probabilities of the respective updated qubits
\begin{eqnarray}
\label{speedup-transport-equation}
p'_+ & \equiv & \langle\psi' | \hat n_1 | \psi'\rangle = p_+ + \Omega_{\hbox{\tiny QLG}}(p_+,p_-)  \\
\nonumber
p'_- & \equiv &  \langle\psi' | \hat n_2 | \psi'\rangle = p_- - \Omega_{\hbox{\tiny QLG}}(p_+,p_-) .
\end{eqnarray}
(\ref{speedup-transport-equation}) implicitly determines the functional form of the mesoscopic collision term $\Omega_{\hbox{\tiny QLG}}(p_+,p_-)$ associated with the microscopic operator $\hat U$.
We use a conservative collision operator as our 2-qubit quantum gate entangling the qubits using only the microscopic states $|01\rangle$ and $|10\rangle$
\begin{equation}
\label{speedup-burgers-general-U2-collision-operator}
\hat U = \left(
\begin{matrix}
1 & 0 & 0 & 0 \cr
0 & e^{i\xi} \cos\theta & e^{i\zeta} \sin\theta & 0\cr
0 & -e^{-i\zeta} \sin\theta & e^{-i\xi} \cos\theta & 0 \cr
0 & 0 & 0 & 1
\end{matrix}
\right).
\end{equation}
 Inserting (\ref{speedup-burgers-general-U2-collision-operator}) into (\ref{speedup-update-equation}), and then substituting the resulting $|\psi'\rangle$ into (\ref{speedup-transport-equation}), we find the collision term $\Omega_{\hbox{\tiny QLG}}$ is \cite{yepez-jstatphy01}
\begin{eqnarray}
\label{speedup-quantum-collision-term}
\Omega_{\hbox{\tiny QLG}} &\equiv& \sin^2\theta (p_- - p_+) \\
\nonumber
&+&  \sin 2\theta\cos(\zeta-\xi)\sqrt{p_+(1-p_+)p_-(1-p_-)}.
\end{eqnarray}
The equilibrium condition $\left.\Omega_{\hbox{\tiny QLG}}\right|_{p=d}=0$ becomes:
\begin{equation}
\label{speedup-burgers-equilibrium-condition}
\frac{d_+}{1-d_+} - \frac{d_-}{1-d_-} = 2\cot\theta\cos(\zeta-\xi)\sqrt{\frac{d_+}{1-d_+}\frac{d_-}{1-d_-}},
\end{equation}
which is a legitimate statement of detailed-balance of collisions at the mesoscopic scale since the quantum model's evolution operator is unitary.
We take the equilibrium occupation probabilities to have the following form:
\begin{equation}
\label{speedup-burgers-fermi-dirac-equilibrum}
d_+  =  \frac{1}{\gamma z + 1}
\hspace{0.25in}
\hbox{and}
\hspace{0.25in}
d_-  =  \frac{1}{\frac{z}{\gamma} + 1}.
\end{equation}
Substituting (\ref{speedup-burgers-fermi-dirac-equilibrum}) into  (\ref{speedup-burgers-equilibrium-condition}) gives a quadratic equation in $\gamma$ that has the solution $\gamma = \sqrt{\alpha^2 + 1} + \alpha$ or $\frac{1}{\gamma} = \sqrt{\alpha^2 + 1} - \alpha$, where $\alpha\equiv\cot\theta\cos(\zeta-\xi)$.  
Next, substituting (\ref{speedup-burgers-fermi-dirac-equilibrum}) into the total number density, $\rho=d_+ + d_-$, we obtain a quadratic equation in $z$
\begin{equation}
\label{speedup-z-quadratic-equation}
\rho z^2+ \left(\gamma + \frac{1}{\gamma}\right)(\rho-1)z + \rho -2 = 0 .
\end{equation}
Substituting the positive root solution of (\ref{speedup-z-quadratic-equation}) into (\ref{speedup-burgers-fermi-dirac-equilibrum}), we find after much algebraic manipulation
\begin{equation}
\label{speedup-d+equilibrium}
d_+ =
\frac{1+\gamma^2 + (1-\gamma^2)\rho -
\gamma\sqrt{\left(\frac{1}{\gamma} + \gamma\right)^2 (\rho-1)^2 + 4(\rho-2)\rho}}
{2(1-\gamma^2)}.
\end{equation}
Then substituting $\gamma = \sqrt{\alpha^2 + 1} + \alpha$ into (\ref{speedup-d+equilibrium}) gives the result
\begin{equation}
d_\pm = \frac{\rho}{2} \mp \frac{1}{2\alpha}\left(  
\sqrt{1+\alpha^2} -\sqrt{1+\alpha^2(\rho-1)^2}
\right).
\end{equation}
This implies that
\begin{equation}
\label{speedup-peqdiff-burgers-qlg}
d_+-d_- = -\frac{1}{\alpha}\left(  
\sqrt{1+\alpha^2} -\sqrt{1+\alpha^2(\rho-1)^2}
\right).
\end{equation}
Again, we compute the components of $J$:
\begin{equation}
J_\pm = \frac{\partial\Omega_{\hbox{\tiny QLG}}}{\partial p_\pm} =  \sin^2\theta 
\left(
\mp1 - \alpha \frac{(2d_\pm -1)d_\mp(1-d_\mp)}{\sqrt{d_+(1-d_+)d_-(1-d_-)}}
\right).
\end{equation}
And this implies
\begin{equation}
\label{speedup-Jdiff-burgers-qlg}
J_+-J_- = -2\sin^2\theta (1+ \alpha^2 f),  
\end{equation}
where the factor $f = f(\alpha, \rho)$ is too complicated an expression to write out here but has the important property that $f(\alpha,\rho) = 1 +{\cal O}(\alpha)$.
Finally, substituting the two results (\ref{speedup-peqdiff-burgers-qlg}) and (\ref{speedup-Jdiff-burgers-qlg}) into (\ref{speedup-effective-field-theory}) gives the effective field theory
\begin{equation}
\label{speedup-burgers-effective-field-theory-qlg-alpha-expanded}
\partial_t\rho+c\cot\theta\cos(\zeta-\xi) (1-\rho)\partial_x\rho=\cot^2\theta\frac{\delta x^2}{\delta t}
\partial_x^2\rho+{\cal O}(\varepsilon^3,\varepsilon\alpha^2),
\end{equation}
which is the nonlinear Burgers equation  for $u=c(\rho - 1)$ with independently tunable sound speed $c_s = 
c\cot\theta\cos(\zeta-\xi) $ and kinematic viscosity $\nu = \cot^2\theta\frac{\delta x^2}{\delta t}$ by appropriately choosing the Euler angles in (\ref{speedup-burgers-general-U2-collision-operator}).
In our quantum case, the trigonometric term $\cot\theta\cos(\zeta-\xi)$ plays the role of the expectation value $\alpha$ of the additional random bit required in the classical model.
Figures~\ref{ViscousQuantumVs3BitVs2Bit} and \ref{QuantumVsViscousClassical}  show the time evolution of the 2-qubit quantum algorithm versus the 3-bit classical Boghosian-Levermore algorithm both carried out on a $N_x=256$ lattice.  The vertical axis is the particle number density $\rho=p_+ + p_-$ plotted in the range of $0.5 \le \rho\le 1.5$.  The time step is in the upper left corner of each snapshot.  The viscosity of the quantum model is close to zero.

The quantum algorithm is unconditionaly stable, obeys detailed-balance, requires less memory than its classical counterpart, and can achieve arbitrarily high Reynolds numbers.  Having a variable transport coefficient that can be made small, it is  consistent with the inviscid Burgers equations when the Euler angle $\theta\simeq \frac{\pi}{2}$.
It is possible to generalize this type of quantum algorithm to three-dimensions to efficiently handle the important application of computational fluid dynamics. 

I would like to acknowledge Professor Boghosian for helpful discussions about his classical lattice-gas model and Owen Cote for helpful discussions about turbulence.

\end{document}